\documentclass[fleqn,10pt,twocolumn]{wlscirep}

\usepackage{pgfplots}
\usepackage[export]{adjustbox}
\usepackage{caption}
\usepackage{subcaption}
\usepackage{lineno}
\usepackage{amsmath}
\usepackage{amssymb}
\usepackage{graphicx}
\usepackage{microtype}
\usepackage{times}

\usepackage[font=footnotesize,labelfont=bf,
   justification=justified,
   format=plain]{caption} 

\pgfplotsset{compat=1.12}

\title{Limitations in Predicting Radiation-Induced Pharmaceutical Instability during Long-Duration Spaceflight}

\author[1,2,*]{Rebecca S. Blue}
\author[3]{Jeffery C. Chancellor}
\author[4,5]{Erik L. Antonsen}
\author[6]{Tina M. Bayuse}
\author[6]{Vernie R. Daniels}
\author[7]{Virginia E. Wotring}

\affil[1]{Aerospace Medicine and Vestibular Research Laboratory, The Mayo Clinic Arizona, Scottsdale, AZ 85054, USA}
\affil[2]{GeoControl Systems, Inc., Houston, TX 77058, USA}
\affil[3]{Department of Physics \& Astronomy, Texas A\& M University, College Station, TX 77843, USA}
\affil[4]{Department of Emergency Medicine and Center for Space Medicine, Baylor College of
Medicine, Houston, TX 77030, USA}
\affil[5]{National Aeronautics and Space Administration (NASA), Johnson Space Center, Houston, TX 77058, USA}
\affil[6]{KBRwyle, Houston, TX 77058, USA}
\affil[7]{Department of Pharmacology and Chemical Biology and Center for Space Medicine, Baylor College of Medicine, Houston, TX 77030, USA}

\affil[*]{Author to whom correspondence should be addressed: Rebecca Blue, rblue.md@gmail.com}


\begin{abstract}
As human spaceflight seeks to expand beyond low-Earth orbit, NASA and its international partners face numerous challenges related to ensuring the safety of their astronauts, including the need to provide a safe and effective pharmacy for long-duration spaceflight. Historical missions have relied upon frequent resupply of onboard pharmaceuticals; as a result, there has been little study into the effects of long-term exposure of pharmaceuticals to the space environment. Of particular concern are the long-term effects of space radiation on drug stability, especially as missions venture away from the protective proximity of the Earth. Here we highlight the risk of space radiation to pharmaceuticals during exploration spaceflight, identifying the limitations of current understanding. We further seek to identify ways in which these limitations could be addressed through dedicated research efforts aimed towards the rapid development of an effective pharmacy for future spaceflight endeavors.
\end{abstract}
\begin{document}

\flushbottom
\maketitle

\section*{Introduction}
With the expansion of human spaceflight outside of low-Earth orbit (LEO), NASA and its international partners face numerous challenges related to ensuring the safety of their astronauts. Among these challenges is the ability to provide a safe and effective pharmacy with sufficient capability to manage both planned and unforeseen medical conditions that may arise during flight. The ability to provide a safe and effective pharmacy to crews is contingent upon multiple factors, such as the stability of any medication for the duration of a given mission, the effectiveness of that medication in the unique space environment, and the provision of appropriate and sufficient medications to meet the unique physiological and psychological challenges the crew may face.

There is a paucity of evidence regarding pharmaceutical stability in the space environment, largely because this issue has not historically been a pressing concern for human spaceflight. Short-duration flights of the Mercury, Gemini, Apollo, and Space Shuttle eras minimized the need for prolonged medication shelf life, and the selection of healthy crewmembers minimized the need for ongoing medication provision for chronic disease. Careful maintenance of crew health and stringent flight rules regarding the more dangerous activities during spaceflight, such as extravehicular activity (EVA), have largely obviated the need for emergency medication provision. Even now, with missions to the International Space Station (ISS) lasting 6 months or longer, crews have been able to rely on medication availability through retirement of expired medications and frequent resupply rather than contending with questions of degradation, storage, and the impact of the space environment (with environmental concerns related to a myriad of factors such as vibration, humidity, and space radiation exposure). As a result, investments in the systematic collection of data for the characterization of medication use, efficacy, side effects, pharmacokinetics, pharmacodynamics, and long-term stability have been a lower priority than other health and human performance investments.  With the push for exploration missions to the moon and Mars, these questions have become a more pressing concern.

One potential risk to pharmaceutical stability arises from long-term exposure to the space radiation environment. While gamma radiation exposure has been used terrestrially for sterilization procedures in select pharmaceuticals, space radiation differs considerably from such practices because of differences in type of radiation, dose, dose-rate, and length of exposure. It is unclear whether long-term exposure to space radiation may affect stability, alter drug ingredients, or produce potentially toxic byproducts, particularly in drugs that have undergone degradation reactions. 

Here, we seek to present the current understanding of pharmaceutical stability in the space radiation environment. In particular, we have attempted to highlight the gaps in current knowledge and the difficulties in translating terrestrial-based radiation studies to a meaningful interpretation of drug response to space radiation. We hope to identify high-yield opportunities for future research that might better define and mitigate the space radiation risk to a future formulary for exploration spaceflight.

\section*{The Interplanetary Space Radiation Environment}
 The effects of radiation are due to the transfer of energy from a charged particle to the medium it travels through.  The amount of energy that can be transferred is a function of the particle's kinetic energy, charge, and mass\cite{attix_introduction_1986,hall_radiobiology_2012}. The effects of indirect ionizing radiation (e.g. gamma, x-ray) are negliglible compared to effects caused by direct ionizing charged particle risk. The more charge a particle has the greater ability it has to ionize the medium it traverses, depositing more energy per unit path length (defined as increased \emph{linear energy transfer}, or LET) in a traversed material.

\begin{figure*}[ht]
\centering
\includegraphics[width=.7\textwidth]{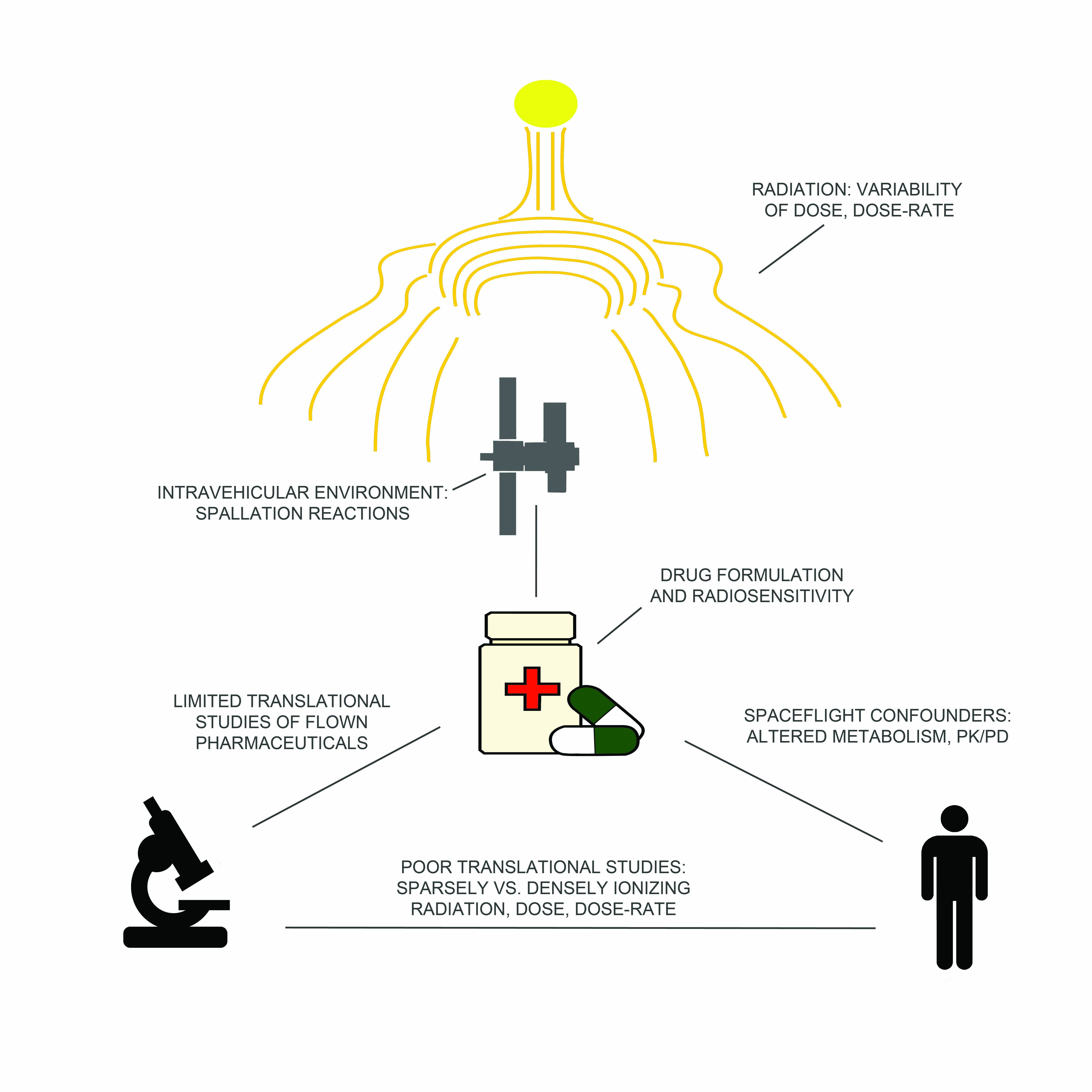}
\vspace*{0.5em}
\caption{Factors limiting understanding of pharmaceutical stability in the space radiation environment. Radiation from galactic cosmic rays (GCR) is not graphically depicted but should be considered ubiquitous in the space environment. PK/PD: Pharmacokinetics/Pharmacodynamics.
}
\label{fig:radpharm-icons}	
\end{figure*}

Future space exploration endeavors will include manned expeditions beyond the protection of the Earth's magnetic field. These long-duration missions, which may span months to years, will require additional protection for the human crews on board. The space radiation environment is a complex mix of charged particles originating from several sources. Within an exploration vehicle (one intended to travel outside of the Earth's geomagnetic field), the the intravehicular radiation environment primarily consists of relativistic heavy-charged particles attributed to galactic cosmic rays (GCR, chronic, isotropic background radiation). The GCR spectrum, and thus the intravehicular radiation environment, primarily consists of ionized hydrogen (protons, approximately 85\%) and helium (alpha, approximately 14\%) nuclei, but also includes less abundant ionized particles of higher atomic weight \cite{ChancellorPosition, zeitlin2013measurements}. Despite their rarity, heavier particles contribute a disproportionately high amount of overall radiation dose-exposure due to their relatively high LET. In addition, as GCR ions pass through vehicular structures, interaction with vehicle materials can cause fragmentation (or "spallation") of heavier ions into more numerous particles of lower atomic weight. This process can produce cascades of ions resulting in a destructive capability in addition to that of the primary ions. Accounting for all such interactions increases the complexity of predicting the intravehicular radiation environment. It is particularly difficult to shield from GCR exposures given the isotropic, highly penetrating nature, and the relative energies of the GCR spectra.\cite{Cucinotta2012Cancer} 

An additional, off-nominal source of charged particle radiation can be attributed to solar particle events (SPEs), where particles are ejected from the sun in prompt and short-lived bursts of energy. SPEs consist primarily of protons and electrons with a relatively small contribution from heavy nuclei. Unlike GCR, SPEs are anisotropic \cite{gombosi1998physics}. SPE radiation is primarily composed of protons with kinetic energies ranging from 10 MeV up to several GeV (determined by the relativistic speed of particles)\cite{ChancellorPosition}. SPEs are capable of accelerating an abundance of protons that can occasionally result in high dose-rates in the interplanetary environment. For example, a particularly large event in October 1989 is predicted to have delivered dose-rates as high as 1,454mGy/hour for a short period of time to an exposed astronaut in a vehicle with 5g/cm$^{2}$ of aluminum-equivalent shielding traveling in interplanetary space\cite{ChancellorPosition,hu_modeling_2009}. While rare, SPE exposure would be in addition to the nominal intravehicular dose, attributed to GCR nuclei, expected to be approximately 0.028mGy/hour during travel in interplanetary space\cite{zeitlin2013measurements}. Interplanetary intravehicular doses would be altered by the peak flux, energy spectrum, and duration of any given SPE as well as shielding thickness and material makeup of the vehicle. Similarly, the contribution of radiation exposure from SPEs and resultant effects on pharmaceuticals would depend upon intravehicular dose and any additional shielding.
%
%


As missions to the moon or Mars will expand human presence from LEO to interplanetary space, intravehicular radiation exposure will increase. Vehicles, and the pharmaceuticals on board, will be exposed to higher cumulative GCR exposure and increased risk for transient SPE exposures. As a result, the risk of radiation-induced alterations of pharmaceutical stability, structure, potency, and potential toxicity will increase with future missions (Figure~\ref{fig:radpharm-icons}).


\section*{Mechanisms of Radiation Impact}

A majority of pharmaceutical radiation risk research is derived from terrestrial analogs rather than the full particle and energy spectrum of the space radiation environment. Accordingly, differences in the relative abilities of terrestrial and space radiation to induce damage in a target have yet to be elucidated. This property of different types of radiation to induce different levels and kinds of damage is known as radiation \textit{quality}. Radiation quality is thought to be dependent on LET, which can be characterized by energy deposition pattern. Charged particles traverse a material in an approximately straight line, transferring energy through interactions with the medium's nuclei and electrons. Imparted energy may be enough to knock an electron out of an atom, ionizing the atom or leaving it in an excited, non-ionized state. The ejected electron can have enough energy to leave the immediate vicinity of the charged particle's path and produce a notable track of its own. This results in a densely ionizing core along the charged particle's path (where energy continues to be deposited in an approximately straight line), as well as a sparsely ionizing penumbra generated by expelled electrons (where energy is deposited throughout the material randomly). 


In addition to differences in radiation quality, substrate composition is an important factor in radiation-induced damage. To date, most radiation research has been conducted in biological models, where a majority of the substrate is water. In this scenario, radiation is more likely to hit water than a biologically relevant target (e.g. DNA). However, even if radiation impacts water rather than a target, it can still induce damage to a target via generation of free radicals, which can diffuse to interact and ionize a target within range. This form of damage is known as \textit{indirect ionization}, while damage caused from radiation hitting a target is known as \textit{direct ionization}. In pharmaceuticals, where a greater percentage of substrate composition consists of target molecules, direct ionization is far more likely than in biological substrates. The difference between percentage of target interactions that are direct versus indirect is highly dependent on substrate and projectile energy.

It has been observed that direct ionization can cause increased damage compared to indirect ionization, particularly in the decomposition of chemical bonds, creation of radiolysis products, and damage to polymer structure.\cite{Kempner2001,Kempner2011,Silindir2009} Studies using biological substrates therefore do not provide good analogues for pharmaceutical research. Furthermore, clustered damage imparted by densely ionizing space radiation combined with the higher target concentration in pharmaceutical formularies could potentially interact, resulting in outcomes that have not yet been characterized.


\section*{Challenges in Reproducing Radiation Dose, Dose-Rate, and Formulation Sensitivity}

Another area of uncertainty is the effect of low doses on pharmaceuticals. Terrestrial pharmaceutical radiosterilization techniques generally use doses of 25-50kGy, which far exceed those expected for even cumulative Mars mission doses (approximately 0.5Gy). Delivery of radiosterilization doses over a matter of minutes or hours considerably exceeds dose-rates anticipated in interplanetary space, where such doses would be accrued over 2-3 years by current estimates. 

It has been suggested that, if a pharmaceutical is found to be stable at higher doses or dose-rates (such as those provided by radiosterilization techniques), then the pharmaceutical should be stable at more limited exposures (such as those delivered in space).\cite{KimPlante}  Some evidence for this argument has been provided by the expected level of damage from indirect ionization. Despite the relatively higher concentration of target molecules in pharmaceutical than biological substrates, damage to pharmaceuticals from indirect ionization does still occur, particularly in liquid pharmaceuticals where target molecules are less concentrated than in solid formulations.  Indirect ionization-induced damage stems from the formation of free radical species; in water-based formulations, this includes the generation of radical oxygen species (oxygen ions and hydrogen peroxide, H$_{2}$O$_{2}$) from the breakdown of water.\cite{Jacobs1977,Jacobs1981,Plessis1979} Studies have indicated that the concentration of these radicals from exposures to radiosterilization doses (25-50kGy) is generally well below toxic levels.\cite{Eisenberg1985,Jacobs2007} A recent NASA technical paper indicated that nanomolar concentrations of radiation byproducts could be produced in exposed pharmaceuticals, but cited low anticipated radiolytic yield in liquid-based pharmaceuticals (based on modeled calculations) as sufficient evidence that irradiated pharmaceuticals should be stable in the space environment.\cite{KimPlante} 

\begin{figure*}[ht]
\centering
\includegraphics[width=\linewidth,keepaspectratio]{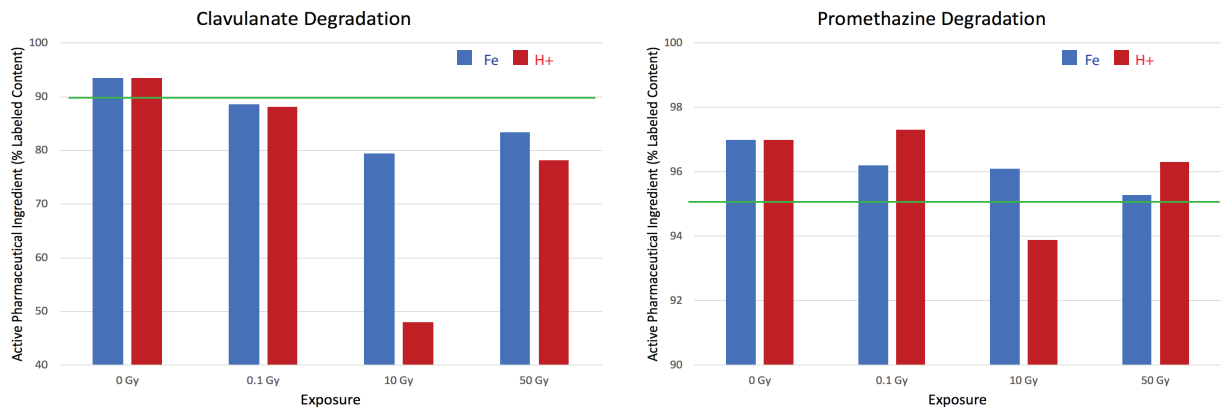}
\vspace*{0.5em}
\caption{NASA data from an NSRL study performed by L. Putcha demonstrating variable drug sensitivity to radiation exposure for clavulanate (as a combination medication, amoxicillin-clavulanate) and promethazine.\cite{Daniels2018} All drug products were measured at time-zero, control and irradiated products were analyzed at the same time following exposures. The solid green line indicates USP-accepted lower limits of percent API content compared to label claims. Note the variable sensitivity both by radiation beam exposure (proton, in red, or iron, in blue) and by dose received (0.1-50Gy). In this study, drugs demonstrated increased degradation to 10Gy exposures compared to 50Gy exposures, suggesting that pharmaceutical stability at higher-dose exposure may not necessarily translate to stability at lower-dose exposures. However, dose and dose-rate of high exposures were significantly greater than even cumulative anticipated doses in long-duration, exploration spaceflight. Further, there is only limited documentation regarding research design or even the full results of this study, limiting our ability to interpret findings.}
\label{fig:brookhaven}
\end{figure*}

However, even the low nanomolar concentrations of ions predicted by that technical report could be enough to sufficiently alter local pH in drug products, which could alter chemistry or drive degradation reactions.\cite{WotringPharm} In addition, studies using electron spin resonance, a sensitive method for the detection of free radicals,\cite{Crucq1996,Gibella2000} have demonstrated that alteration of radiation dose changes the concentration and type of free radicals produced, often with unpredicted complexity or type of resultant radical species. \cite{Gibella2000} These complex reactions could alter subsequent radical-induced damage.\cite{Gibella2000} Dose-rate may be an important factor in the activity of radical species, and many pharmaceuticals are demonstrated to be more stable at higher dose-rates. It has been theorized that high dose-rate increases oxygen consumption, resulting in decreased presence of oxygen radicals (or the rapid consumption of any radical species generated) and associated damage.\cite{Gibella2000,Jacobs2007} Shorter-duration exposures may produce fewer long-lived oxygen radical species and, as a result, less prolonged opportunity for delayed damage than protracted exposures. In evidence to these arguments, one historical study of spaceflight-approved pharmaceuticals compared drug stability at variable radiation dose ranging from 0.1-50Gy and found drug degradation associated with moderate radiation exposure where no instability was noted at higher doses.\cite{Daniels2018} It is worth reiterating that these dose ranges include exposures that are substantially greater than even cumulative anticipated doses in long-duration, exploration spaceflight.

Liquid pharmaceuticals are often considered less stable than solid or powdered drugs given the greater potential for free radical formation in water-based formulations, the possibility of interactions between substrate and \textit{excipients} (the pharmacologically inert compounds in a given dosage formulation), incomplete dissolution of substrate, crystallization of dissolved compounds, and other alterations of drug suspensions over time. Water-based drugs will undergo more frequent hydrolysis reactions, driving more prevalent and more rapid degradation reactions. As a result, the rare discussions of pharmaceutical stability in the context of space radiation have focused on liquid formulations and postulated that, should liquid formulations be determined to be stable in the space radiation environment, solid or semi-solid formulations would be of no additional concern.\cite{KimPlante} However, some studies demonstrate radiation-induced instability in solid or powder formulations, with reports of radical trapping in excipient lattices leading to a longer presence of free radicals in powder or solid drugs than in liquid formulations.\cite{ambroEPR,Dicle2015,Koseoglu2003,Taiwo1999} In addition, the interaction of particles with solid or powder substrates may produce increased types and complexity of ion species due to spallation. Spallation ions impacting stored pharmaceuticals may cause increased direct and indirect ionizations or induce additional chemical reactivity in the substrate. In short, it is unclear how drugs of any formulation may respond to the unique qualities of space radiation, simply because such responses have not been studied to any degree of fidelity.

Given the uncertainty of drug response to alterations of dose, dose-rate, or exposure time, radiosterilization is only approved for well-documented procedures of declared dosage (most commonly 25-50kGy) and dose-rate, and deviation from the designated dose or dose-rate is assumed to be capable of altering the final drug product.\cite{Gibella2000,Jacobs2007} In evidence to this concern, the United States Pharmacopeia (USP) regards radiosterilized pharmaceuticals as entirely new products, and pharmaceutical companies are required to submit new drug applications and demonstrate safety, potency, and lack of toxic breakdown products for approval of radiosterilization in any marketed drug.\cite{Jacobs2007,USPharmacopeia} The use of terrestrial analog radiation to predict the response of pharmaceuticals in the space environment directly contradicts the standard approach to safety and stability review of irradiated pharmaceuticals. Finally, it should be noted that many of the pharmaceuticals currently included in a spaceflight formulary are not approved by the U.S. Food and Drug Administration (FDA) for terrestrial radiosterilization procedures.

\section*{Challenges in Emulating the Space Environment}
Accurate simulation of the complex space radiation environment for pharmaceutical testing via terrestrial analog is currently not possible, given limitations in radiation type and dose-rate of exposure. Space radiation studies, pharmaceutical or otherwise, often make use of a recently updated GCR simulator at the NASA Space Radiation Laboratory (NSRL) at Brookhaven National Laboratory in Brookhaven, New York. To date, the NSRL is the only U.S. government facility capable of generating heavy-charged particles at energies and spectra that approximate the space environment.\cite{norbury_galactic_2016,Slaba2015} Recent improvements now allow for rapid switching between ion species, providing rapid and consecutive exposures to different mono-energetic ion beams.\cite{norbury_galactic_2016} Rapid switching of ion beams may be sufficient in simulation of the complex space environment, particularly as previous modeling has suggested that the likelihood of multiple ion species traversing a small volume at the same time (i.e. traversing a single drug tablet) is exceedingly low.\cite{Curtis2013} This rapid switching technique is a potential improvement when compared to use of photon or single ion exposures and offers more insight than studies with considerably higher doses than those expected during spaceflight, such as  the doses used in radiosterilization literature.\cite{ChancellorPosition,norbury_galactic_2016}

However, even this simulator utilizes exposures that are appreciably different from those anticipated in the interplanetary space environment, generally delivering cumulative anticipated mission doses over short periods of time.\cite{norbury_galactic_2016} While the NSRL simulator is capable of providing more protracted doses, limitations of funding for long-term experiments generally limit exposure times, causing deviation of the analog from GCR. The simulator cannot generate the full spectrum of ions or spallation ions that make up the GCR spectrum; instead, exposures are limited to only a sampling of some of the heavy ions that contribute to GCR, and these ions are delivered sequentially rather than simultaneously. Further, the simulator lacks the capacity to generate the pions (subatomic particles) or neutrons that would follow spallation reactions in the intravehicular environment,\cite{ChancellorPosition,Slaba2015} though these would be expected to account for 15-20\% of an intravehicular exposure.\cite{chancellor_emulation_2017,Slaba2015}  These factors may limit ability to translate terrestrial analog studies to an understanding of the true risk of space radiation pharmaceutical exposure.

Even so, the NSRL simulator is one of the few simulators available to study space-like radiation in the terrestrial environment. To date, there are remarkably few studies of pharmaceuticals at this facility. In 2011, Chuong et al. studied the stability of solid formulations of vitamin B during spaceflight and in terrestrial radiation analogs, making use of an older radiation simulator at the NSRL for some exposures.\cite{Chuong2011} The authors studied vitamins that had flown onboard the Space Shuttle and ISS for 2-4 weeks or 12-19 months, comparing them to terrestrial controls and vitamins exposed to the terrestrial radiation beam. While the NSRL exposures were used as a terrestrial radiation study arm to examine radiation effects on the vitamin, it is noteworthy that the NSRL exposures were mono-energetic exposures of 0.1-50Gy, using either hydrogen or iron radiation sources.\cite{Chuong2011} The upper limits of exposures in this study greatly exceed those expected during even long duration and exploration spaceflight. 

The USP allows variation of the \textit{active pharmaceutical ingredient }(API, the ingredient imparting the desired physiological effect of a medication) within 90-150\% of package label content for vitamin B. The Chuong study did identify statistically significant, though acceptable, variation in content of ground (unflown, non-irradiated) and NSRL irradiated samples, and even identified one flown sample with API concentration well below acceptable ranges.\cite{Chuong2011} However, the authors suggest that instability was most likely related to API formulation, excipient interaction, or even packaging, and stated that, since NSRL samples were found to be stable (at notably higher dose-rate than flown samples), radiation was not the cause of instability.\cite{Chuong2011} As discussed above, this reasoning is questionable given the numerous factors that limit translation of simulated radiation exposures to the true space environment. 

Additional research conducted at the NSRL by NASA on numerous pharmaceuticals in an effort to delineate the effects of various radiation doses on drug stability similarly utilized 0.1-50Gy doses of proton or iron mono-energetic beams.\cite{Daniels2018} Data released suggests dose-variable alterations of API, with increased degradation noted at 10Gy exposures compared with those at 50Gy exposures delivered over equivalent time intervals (Figure~\ref{fig:brookhaven}).\cite{Daniels2018} Study exposures are still notably higher than those expected during spaceflight but suggest that stability at high-dose exposures may not necessarily translate to stability at low-dose exposures, and that dose and dose-rate alterations may significantly impact stability. Unfortunately, there is only limited documentation regarding research design or even the full results of this study, limiting our ability to fully interpret findings.

\section*{Mechanisms of Pharmaceutical Instability}
Pharmaceuticals can become unstable through alteration of either their physical or chemical properties. Alteration of physical properties includes changes in appearance or consistency; alteration of chemical properties includes loss of potency, alteration of excipients, excipient-active ingredient interactions, or toxic degradation.\cite{mehta2017,RhodesPharm} In order to determine that a pharmaceutical is unchanged by exposure to the radiation environment, a drug must be demonstrated following exposure to have no significant alteration of its API(s) while at the same time have no significant development of degradation products that are either toxic themselves or in some way alter the pharmaceutical properties of the original medication.\cite{WotringPharm} The USP provides guidelines for acceptable API content in medications approved by the FDA, commonly within 10\% of label-specified content (though this can vary considerably by drug type or API).\cite{USPharmacopeia} A medication would be considered radiosensitive if API concentration fails to meet USP requirements following radiation exposure. Alterations of API can affect drug potency, efficacy, and safety rendering the drug less effective, ineffective, or potentially dangerous. 

There are numerous documented cases of pharmaceuticals being altered by radiation exposure at sterilization doses (25-50kGy). For example, irradiation of metoclopramide hydrochloride produced a number of degradation products following radiation exposure,\cite{Maquille2008} and gamma sterilization of certain beta-blockers has been demonstrated to alter the pharmaceuticals' color and appearance and affect the melting point of the drug preparations.\cite{Marciniec2011} Even compounds that are molecularly similar may have vastly different responses to irradiation.\cite{Hasanain2014} Cephradine and cefotaxime, both solid-form cephalosporin antibiotics of similar molecular structure, demonstrate significantly different radiosensitivity when exposed to identical sterilization doses of gamma radiation. Cephradine degrades significantly and has been determined to be unstable under irradiation\cite{Signoretti1993} where cefotaxime demonstrates high resistance and stability.\cite{Barbarin1996,Barbarin2001}  As molecular alterations can change the saturation of the compound or the presence or absence of reactive groups such as alcohols, acids, or ketones, even minor differences in API structure can affect radiosensitivity.

In addition to altering API, radiation exposure can result in the generation of degradation products and may alter the medication, whether or not the API is affected, by damaging the structure or action of excipients. For example, radiation is known to alter the chemical structure of various polymer drug delivery systems, causing increased cross-linking of polymers in some cases and inducing polymer chain breakage in others.\cite{Chapiro1974,Hasanain2014} Cross-linked polymers, with higher molecular weight, may cause issues with insolubility,\cite{Martini1997} and chain breakage of some polymeric microspheres used for drug delivery have been associated with high production of free radicals and instability of the resultant compounds.\cite{Hasanain2014,Montanari1998} In some studies, alteration of excipients has been demonstrated to affect dissolution rates and controlled release of API.\cite{Maggi2003,Maggi2004} 

Radiosensitivity is highly specific to dose, dose-rate, radiation type (photon, electron, proton, heavy ion, etc), chemical composition, excipient content, and drug formulation. However, it must also be stated that any radiation-induced pharmaceutical risk must be weighed in the context of the multitude of other factors that may render a flown drug unstable in the space environment. Mission duration will soon extend beyond approved shelf life for many medications currently included in onboard medical kits. Current medications aboard the ISS are replenished through regular resupply and removal of older drugs; this may not be possible with future missions to the moon or Mars.\cite{WotringPharm}  Older drugs may be at higher risk of degradation from chronic exposure to the radiation environment. 

NASA currently repackages some of the flown pharmaceuticals to manage mass and volume constraints and to limit packaging waste in the closed environment of a space vehicle. However, repackaging itself may affect shelf life or stability of stored medications or alter their response to radiation exposure.\cite{WotringRisk2011,WotringPharm} For example, nuclei interacting with packaging material could produce additional progeny ions that alter the chemical composition of pharmaceuticals within.\cite{mehta2017} Previous studies have suggested various packaging materials that may be intrinsically better for radiation shielding, such as polyethylene;\cite{mehta2017,Guetersloh2006,Wook2014,Harrison2008} however, there are insufficient data regarding ideal packaging technique or long-term shelf life of pharmaceuticals packaged in such materials, and any novel packaging approach intended for use onboard future missions would be subject to USP review and guidelines.\cite{USPharmacopeia} Ultimately, choice of packaging materials should address radiation sensitivity as well as additional shelf life concerns, particularly as it remains unclear how the factors of drug age, repackaging, shelf life, and radiation exposure will interact to determine pharmaceutical response.

\begin{figure*}[ht]
\centering
\includegraphics[width=\linewidth,keepaspectratio]{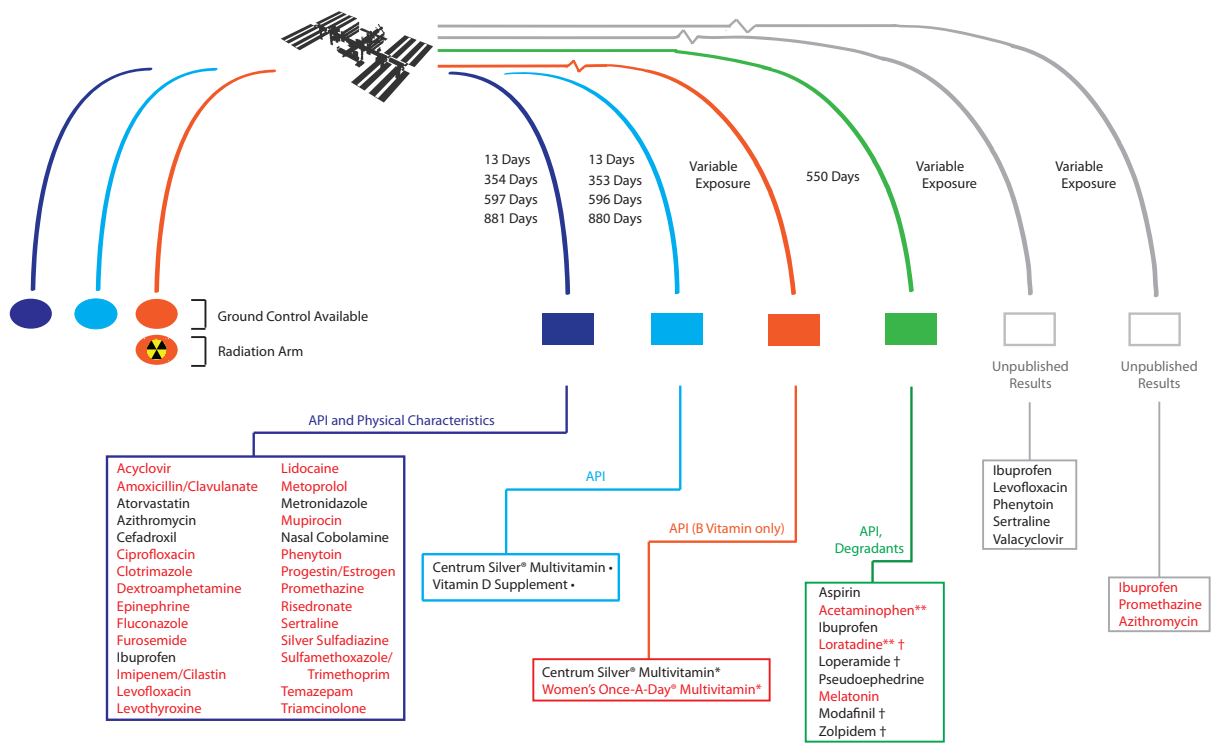}
\vspace*{0.5em}
\caption{To date, there have been few studies of pharmaceuticals flown in the space environment. The studies presented in the figure included various evaluations of active pharmaceutical ingredient (API), physical characteristics, impurity products, and degradation, as indicated \cite{Wotring2016, du_evaluation_2011, Zwart2009, Chuong2011, Cory2017, Wu2016}. Only one study by Chuong et. al.\cite{Chuong2011} included "radiation arm," a subset of ground controls that were irradiated with either hydrogen or iron ions at high dose and dose-rate dissimilar to the space environment. Drugs in red text were found to have alterations of API, physical characteristics, or contain significant concentrations of degradants or impurities after flight in one or more preparation of the indicated pharmaceutical. *Multivitamin preparations were analyzed only for B-complex API stability. **Drugs contained API concentrations within acceptable limits at time of study analysis, but would fail API analysis according to current standards. \textdagger Drugs contained unspecified or unidentified impurity products of unknown significance. \textbullet  Multivitamin content demonstrated time-related instability but showed no alteration specifically related to spaceflight exposure.}
\label{fig:flownstudies}
\end{figure*}

There have been very few examinations of pharmaceuticals actually exposed to the space environment, including a ground-controlled study of API in flown pharmaceuticals conducted by Du et. al.\cite{du_evaluation_2011} and a convenience sampling of pharmaceuticals returned to Earth after >550 days aboard the ISS by Wotring (see Figure~\ref{fig:flownstudies}).\cite{Wotring2016} More recently, Cory et. al.\cite{Cory2017} and Wu et. al.\cite{Wu2016} sought to analyze potency, purity, and drug degradation in certain pharmaceuticals flown aboard the ISS, though these results have yet to be published. Two additional papers addressed multivitamin stability after spaceflight exposure, including Chuong et al.\cite{Chuong2011} and Zwart et al.\cite{Zwart2009}. In general, most pharmaceuticals tested after flight have been found to meet USP requirements for API concentration, though notable exceptions occurred. For example, Du et. al. found that amoxicillin-clavulanate, levofloxacin, trimethoprim, sulfamethoxaxole, furosemide, and levothyroxine degraded before their expiration dates.\cite{du_evaluation_2011,WotringPharm} The study additionally identified alterations of physical appearance of some medications.
Wotring identified degradation and impurity products in aspirin, ibuprofen, loratadine, modafinil, and zolpidem.\cite{Wotring2016} The two multivitamin studies identified alteration of multivitamins over time in both ground and flown samples when compared to time-zero controls, but neither found convincing evidence of degradation specific to spaceflight-flown formulations\cite{Chuong2011,Zwart2009}. While these studies have provided at least some much-needed pilot data, they are limited by the ability to provide adequate ground-control, control of confounders, or appropriate reproducibility given limited sample size, and can provide only an initial awareness that flown pharmaceuticals may not be stable in the space environment. It is worth emphasizing that pilot data do suggest that expected radiation exposures may be sufficient to affect medication stability. While yet unpublished, reported results from more recent experiments performed by Cory et. al. and Wu et. al. were similarly limited by exposure-time variables, limited ground controls, and drug lot variability, but again suggest instability despite theoretical expectations to the contrary.\cite{Cory2017,Wu2016} 

Finally, despite decades of pharmaceutical use in spaceflight, there is limited knowledge regarding alterations of pharmacokinetics (absorption, metabolism, and excretion of a medication) and pharmacodynamics (drug effects on the body) in the space environment. As the human body undergoes significant physiological and metabolic changes during spaceflight, it stands to reason that the effects of pharmaceuticals on an astronaut may change during flight.\cite{Gandia2005} However, research on this issue has largely been limited to observational reports and analog studies.\cite{WotringPharm,Gandia2005} Without directed studies to examine the multifactorial impact of the space environment on pharmaceutical response, it is difficult to fully understand how the additional risks from space radiation may further alter drug response, if at all, during exploration missions.

\section*{Discussion}

Numerous confounders, limited spaceflight studies, and challenges in translation of terrestrial analog evidence to spaceflight have all hindered our ability to draw meaningful conclusions regarding the stability of pharmaceuticals during exploration spaceflight. As NASA looks towards the challenges associated with missions involving increased distance from Earth, the current inability to provide a safe and effective pharmacy for exploration spaceflight has been identified as a major research gap.\cite{AntonsenRISK} To address this issue, NASA recently developed a Pharmacy Research Plan in which pharmaceutical stability and radiation risk are highlighted as unknowns that should be addressed in dedicated research efforts prior to lunar or Mars missions.\cite{Daniels2017} However, this research plan faces challenges including approaching mission design-freeze deadlines and a need to declare a planned formulary for fast-approaching exploration missions, expected to occur within the next decade of spaceflight.

As an adjunct to NASA's research plan, recent literature has provided potential solutions for storage- and radiation-related stability concerns. For example, there has been some suggestion that cryogenic storage conditions may be protective to pharmaceuticals during spaceflight.\cite{mehta2017,Jaworske2016} Such methods have been demonstrated to be successful during radiosterilization processes, providing increased stability of medications during gamma or x-ray exposure.\cite{mehta2017,Meents2010,Moyne2002} Even so, some formulations may demonstrate decreased stability with freezing; for many drugs, effects are unknown or unstudied. There have been no studies of cryogenically stored pharmaceuticals exposed to space-like radiation doses, dose-rates, or spectral complexity. It is difficult to predict the response of cryogenic pharmaceuticals to the space environment, given the multitude of confounding factors and the paucity of data available. 

Similarly, previous literature has discussed the potential inclusion of "space-hardy" formulations, such as use of excipients believed to be more stable in a radiation environment.\cite{mehta2017,Wotring2016} For example, formulations including starch, stearate, cellulose, and dextrose may be more likely to be stable than alternatives, based on results from the 2016 Wotring study\cite{mehta2017,Wotring2016} Other options include preparations including excipients such as mannitol, nicotinamide, and pyridoxine, which have demonstrated radioprotective properties in terrestrial sterilization processing.\cite{mehta2017} A more thorough discussion of potential excipients for improved stability, radioprotective qualities, and antioxidant effects can be found in Mehta et al.\cite{mehta2017} However, it should be reiterated that much of the literature supporting inclusion or exclusion of excipients for protective or stability properties is again based on incomplete data, convenience sampling, or radiation exposures dissimilar to the space environment, limiting the translation of findings particularly for long-duration, exploration missions. Further, altering or adding excipients would change drug formulation; the resultant product would be considered a new drug and, per USP regulations, would require a new application for drug approval and demonstration of safety, potency, and lack of toxic breakdown products.

Finally, there has been discussion of limiting the impact of pharmaceutical irradiation through the inclusion of onboard shielding.\cite{Jaworske2016} In collaborative efforts to protect human crew from radiation exposure, there has been much discussion regarding the inclusion of a heavily-shielded compartment of thick aluminum or other radioprotective material, or alternatively by the use of "multi-purpose shielding solutions," such as barriers composed of water or food supplies, on exploration vehicles.\cite{Jaworske2016,huff2016evidence,Simon2017Shield} Pharmaceuticals could be stored within a shielded compartment to reduce radiation exposures. While these innovative efforts show promise, it is important to remember that shielding designs may be limited by mass and volume constraints and lift-mass capabilities of the launching vehicle. It is premature to assume that idealized shielding will be successfully implemented in early exploration vehicles. Many of the shield designs are intended for "just-in-time" deployment for protection of crew\cite{Simon2017Shield}; in such circumstances, crew would have to retrieve onboard pharmacy stores and transfer them into the shielded space to protect drugs from SPEs. Even if a high degree of shielding were to be implemented, such a compartment would only mitigate transient exposures associated with large SPEs, and protracted exposure to GCR would continue to pose a threat to drug stability in long-duration spaceflight. 



Ultimately, successful mitigation of radiation risk relies upon a more thorough understanding of the potential effects of radiation upon pharmaceuticals, insight regarding which pharmaceuticals are at highest risk for radiation-induced damage, and an awareness of how the myriad of spaceflight-related factors (e.g. altered pharmacokinetics and pharmacodynamics, radiation dose, radiation dose-rate, packaging, shelf life, etc.) affect an exposed drug. Careful and controlled study of pharmaceutical stability, with ground controls and appropriate sample size, would greatly improve our understanding of the multifactorial risks to pharmaceuticals in space. Additional ground-based studies comparing the effects of gamma, x-ray, or electron beam to proton or heavy ion exposure may improve understanding of how to better translate terrestrial literature to the context of space radiation. Utilization of the ISS as a research platform, with long-duration storage of pharmaceuticals and well-designed and controlled studies of shelf life and radiation exposure, could provide much-needed understanding of stability in actual spaceflight conditions. However, such studies would need to be initiated rapidly, as the ISS is intended for decommissioning within the next decade. With rapidly approaching exploration mission dates, NASA and its international partners seek a mature pharmaceutical formulary that can be realized before vehicle and mission design freezes occur. Inclusion of pharmaceuticals (particularly novel or complex pharmaceuticals not currently included in the ISS formulary) onboard future manned or unmanned missions outside of LEO could provide additional data of drug stability in the actual deep-space environment; however, most of these missions do not return payloads to the Earth, limiting analysis options. Maximizing analysis opportunities requires programmatic commitments to sample return; this in turn depends upon an increased understanding of the importance of these data. 

As a comprehensive research plan onboard the ISS may not be feasible given time and financial constraints, study of comparative effects of single- and multi-energetic exposures may improve our understanding of the complexity of the space radiation environment and its impact on pharmaceuticals. As discussed above, use of dose and dose-rates that more closely emulate the space environment may provide more useful or accurate results than reliance upon high dose and dose-rate exposures. Further, inclusion of shielding materials in terrestrial analog design, to both mitigate dose and simulate potential spallation reactions, may better mimic the ionic composition of the intravehicular environment.\cite{chancellor_emulation_2017} Comparative studies of drugs both including and excluding shielding or packaging materials may provide insight regarding the relative contribution of such materials to pharmaceutical degradation reactions.

Careful and thorough evaluation of pharmaceuticals exposed to the space environment is overdue, and the paucity of data limits appropriate translation of terrestrial studies for understanding of space radiation exposures. It is critical to address these knowledge gaps before missions to the moon or Mars are underway. There are significant advances that can be achieved by a well-planned research effort that provides both actual flight data from flown pharmaceuticals aboard available research platforms, such as the ISS, as well as translational studies of comparative effects of space-like radiation in terrestrial analogs, allowing for better interpretation of historical terrestrial radiation understanding in the context of spaceflight. Use of improved and robust modeling techniques that better emulate the space environment, careful study of various formulations, alternate drug choices, and packaging materials, and consideration of novel techniques, such as cryogenic storage, could provide much-needed advances towards the development of a pharmaceutical capability for interplanetary flight. 


\vspace*{0.5em}

\section*{Acknowledgments}

\noindent
The authors acknowledge Kerry Lee, Caitlin Milder, and S. Robin Elgart, in association with NASA's Space Radiation Analysis Group, for their assistance. We further acknowledge the support of the Exploration Medical Capability Element and the Clinical Pharmacy at NASA Johnson Space Center.  The views and conclusions contained herein are those of the authors and should not be interpreted as necessarily representing the official policies or endorsements, either expressed or implied, of the U.S.~Government. The U.S. Government is authorized to reproduce and distribute reprints for Governmental purpose notwithstanding any copyright annotation therein. JC acknowledges the Texas Advanced Computing Center (TACC) at The University of Texas at Austin for providing HPC resources that have contributed to the research results reported within this paper.

\section*{Competing interests}

\noindent The authors declare that they have no competing interests.

\section*{Author Contributions} \noindent
RB developed the concept of the review. JC contributed to the discussion on space physics. All authors contributed to the review of the literature, discussion on the interpretation of research outcomes to spaceflight operations, and drafting of the manuscript.

\end{document}